\documentclass[aps,floatfix,prb]{revtex4}
\usepackage{graphicx,amsmath,amsfonts,amssymb,enumerate,color}

\makeatletter
\newcommand\ket[1]{|#1\rangle}

\begin{document}

\title[Spin fluctuations in a noisy magnetic field]{Creation and dynamics of spin fluctuations in a noisy magnetic field}

\author{J. Delpy $^1$, S. Liu$^{1,2}$, P. Neveu$^{1}$, C. Roussy$^{1}$, Th. Jolicoeur$^{3}$, F. Bretenaker$^{1}$, F. Goldfarb$^{1,4}$ }

\affiliation{$^1$ Universit\'e Paris-Saclay, CNRS, Ecole Normale Sup\'erieure Paris-Saclay, CentraleSup\'elec, LuMIn, Orsay, France}
\affiliation{$^{2}$ East China Normal University, State Key Laboratory of Precision Spectroscopy, Shanghai, China}
\affiliation{$^{3}$ Universit\'e Paris-Saclay, CNRS, CEA, Institut de Physique Th\'eorique, France}
\affiliation{$^4$ Institut Universitaire de France (IUF)}

\begin{abstract}

We theoretically and numerically investigate the spin fluctuations induced in a thermal atomic ensemble by an external fluctuating uniaxial magnetic field, in the context of a standard spin noise spectroscopy (SNS) experiment. We show that additional spin noise is excited, which dramatically depends on the magnetic noise variance and bandwidth, as well as on the power of the probe light and its polarization direction. We develop an analytical perturbative model proving that this spin noise first emerges from the residual optical pumping in the medium, which is then converted into spin fluctuations by the magnetic noise and eventually detected using SNS. The system studied is a spin-1 system, which thus shows both Faraday rotation and ellipticity noises induced by the random magnetic fluctuations. The analytical model gives results in perfect agreement with the numerical simulations, with potential applications in future experimental characterization of stray field properties and their influence on spin dynamics.

\end{abstract}

\maketitle

\section*{Introduction}

Coherent control and stabilization of spin states is an ubiquitous challenge in the field of quantum technologies, including quantum information or sensing. Indeed, the interaction of a spin with its nearby environment is unavoidable and eventually leads to the mixing of the quantum states of the spin and its bath \cite{szankowski_environmental_2017}. Interestingly, this coupling can be characterized using techniques such as spin noise spectroscopy (SNS) \cite{aleksandrov1981magnetic}, which can be used to measure the spontaneous stochastic fluctuations of an ensemble of spins and the associated decoherence mechanisms. Such experiments were conducted in condensed structures to probe electron-nuclei spin interactions \cite{poltavtsev_spin_2014}, confinement effects \cite{muller_spin_2008} or electron-hole coupling \cite{crooker_spin_2010}. 
In thermal vapors, relaxation processes such as spin exchange in a single specie \cite{katsoprinakis2007measurement, Dellis_Loulakis_Kominis_2014} or two-species atomic samples \cite{mouloudakis_quantum_2019, roy2015cross}, as well as binary collisions \cite{Song_Jiang_Qin_Tong_Zhang_Qin_Liu_Peng_2022} or simply atomic motion \cite{Liu_2022} were studied.

It is well know that such spin fluctuations are sensitive to external fields. The induced coupling can be used to probe non-linear SNS regimes, with respect to magnetic field \cite{Mihaila_Crooker_Rickel_Blagoev_Littlewood_Smith_2006}, with application to magnetometry \cite{swar_measurements_2018}, or with respect to light fields \cite{PhysRevA.103.023104}. Although first used as a non-invasive technique, intense probe beams or resonant driving fields were proved to be useful to reveal non-equilibrium features \cite{li_nonequilibrium_2013, sinitsyn_theory_2016}, such as ground state coherences \cite{glasenapp_spin_2014} or optical coherences \cite{PhysRevA.107.L011701, horn_spin-noise_2011_}. Moreover, it has been demonstrated recently that an ac magnetic field could be responsible for higher-order spin noise correlation: 4th order correlators were shown to carry information on the coupling of all spins in the ensemble due to the oscillating field \cite{li_higher-order_2016}. In most cases however, additional decoherence follows \cite{szankowski_spin_2013, li_higher-order_2016}. In the worst case, random fluctuating field such as stray fields can degrade the acquired spin noise spectra without beneficial counterparts.

The effect of random magnetic fields is even more crucial when considering the case of optically pumped magnetometers (OPM). Such devices rely on the measurement of the Faraday rotation angle of a probe light interacting with an optically pumped atomic vapor \cite{budker_optical_2007, kominis_subfemtotesla_2003}. In this case, the value of a magnetic field is inferred by measuring the Larmor frequency at which the spins precess. The sensitivity of these devices strongly depends on the thermal spin noise of the ensemble \cite{shah_high_2010}, with a lower limit set by purely quantum constraints on the minimimum spin projection noise \cite{vershovskii_projection_2020, wolfgramm_squeezed-light_2010, koschorreck_sub-projection-noise_2010}. On the other hand, the impact of magnetic noise on OPM measurement has been pointed \cite{kim_ultra-sensitive_2016,wilson_ultrastable_2019}, and a magnetic noise stabilization scheme based on dual-species cells have been recently proposed \cite{ding_dual-species_2023}. However, the link between the spin noise in the ensemble and the magnetic noise itself is yet to be clarified.

In this paper, we theoretically and numerically investigate the effect of a uniaxial noisy magnetic field on the spin dynamics in a thermal atomic ensemble, within the framework of a standard SNS experiment. In section 1, we first recall the spin noise spectroscopy experiments principles, and compare some experimental results conducted in metastable Helium, near a $J=1\rightarrow J=0$ ($\mathrm{D_0}$) transition, both in a clean and magnetically disturbed environment. We show the impact of a magnetic noise on Faraday rotation (FR) noise power spectra. 

In section 2, based on a microscopic model for such  an atomic transition, we investigate numerically whether such a randomly fluctuating field can create additional spin fluctuations with noise levels comparable to the one observed in motion-limited spin noise conditions.

In section 3, we implement a perturbative treatment to support our numerical simulation results with analytical results, and to question the physical origin of this noise. Using the decomposition of the spin oscillations in eight degrees of freedom corresponding to different spin arrangements \cite{colangelo_quantum_2013, PhysRevA.107.023527}, we study the impact of the statistical properties of the magnetic noise. The question then is to understand how the excitation of the spin fluctuations depends on the central frequency and bandwidth of the magnetic noise. We then try to gain further physical insight into the creation of the spin noise, by investigating the impact of the steady-state around which the fluctuations occur.

The last section discusses  higher-order tensor spin noise, probed as ellipticity noise in a standard SNS setup. Such a noise has already been studied as a consequence of transit noise in high spin systems \cite{fomin_spin-alignment_2020, Liu_2022}. After having studied the creation of circular birefringence noise, we naturally further investigate the possible existence of linear birefringence fluctuations induced by the fluctuating magnetic field.

\section{Motivation : biased experimental results}

\subsection{Principle of spin noise spectroscopy}

\begin{figure}[h]
\includegraphics[width=\columnwidth]{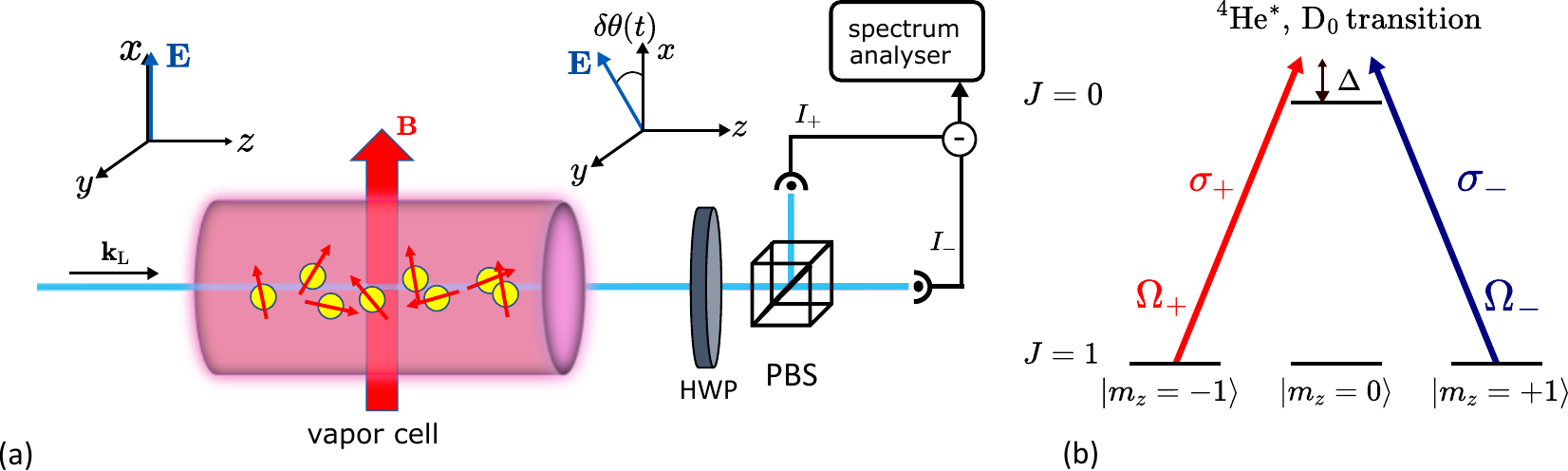}
\caption{(a) Schematics of a standard SNS experiment in a thermal vapor~: 
a linearly polarized probe beam interacts with a paramagnetic atomic sample.
The fluctuations of the average spin in the volume results in a stochastic
Faraday-like rotation of the polarization by an angle $\delta \theta(t)$. 
(b) Energy diagram for a $J=1 \rightarrow J=0$ transition, with Zeeman
sublevels quantized in the direction of the probe wavevector 
$\mathbf{k_\mathrm{L}}$. In this basis, the states $\ket{m_z = \pm 1}$ 
are coupled to the excited level by the $\sigma_\pm$ polarization states 
of  light, with Rabi frequencies $\Omega_\pm$. The light is detuned by 
a frequency $\Delta$ from the center of the transition.}
\label{experiment_diagram}
\end{figure}

Let us first recall  the very basics of spin noise spectroscopy  (SNS) experiments. The schematics of a standard SNS setup is presented in figure \ref{experiment_diagram}(a). By sending a linearly polarized beam through a sample of interest, the fluctuations of the total spin contained in the laser volume are probed optically. To do so, one measures the stochastic Faraday rotation experienced by the probe polarization, created by the fluctuations of the projection of the spin along the light propagation axis. Since these fluctuations are responsible for circular birefringence noise, using a perfectly linearly polarized probe light maximizes the Faraday rotation effect, and thus ensures the optimal detection of the noise. Small ellipticity defects in the polarization are negligible in first approximation but would degrade the measurements if too large. The induced tiny angles of rotation are then measured using a balanced detection. A polarizing beam splitter (PBS) separates the light into two beams with orthogonal polarizations, which are sent on two photodiodes. Consequently, the stochastic rotation noise induces fluctuations of the intensities $I_{\pm}$ measured by the two photodiodes. The photocurrents are then subtracted. The remaining current is amplified and fed into an electronic spectrum analyser to get the spin noise power spectral density (PSD). To shift the spin noise resonance out of the frequencies where laser and electronic noises dominate, a dc transverse magnetic field is applied, with a magnitude of a few tenth of Gauss to a few Gauss. This magnetic field centers the spin noise resonance around the Larmor frequency $\omega_\mathrm{L}$, in the range of hundreds of kHz to a few MHz, higher than other technical noises. Moreover, the use of a balanced detection helps suppressing other additional perturbations such as the laser intensity noise. As a consequence, this experimental setup allows for the optical measurement of the intrinsic dynamics of a spin ensemble in an external magnetic field. 

\subsection{Former experimental results in presence of stray magnetic fields}
The motivation for the theoretical study presented in this paper is the differences between recent and older SNS results, recorded in two different laboratory facilities for the same system.

We conducted SNS experiments in a thermal vapor of metastable Helium ($^4\mathrm{He}^*$). A $6\,\mathrm{cm}$-long cell is filled with Helium atoms at a pressure of 1 Torr. A radiofrequency discharge at $27\, \mathrm{MHz}$ creates a plasma, in which collisions bring a fraction of the atoms from the ground state to the $\ket{2^3S_1}$ metastable state leading to a density of metastable atoms in the cell of the order $10^{11}\,\mathrm{cm^{-3}}$. We probe the spin fluctuations of these excited atoms using a fiber laser with a diameter reduced to $0.6\,\mathrm{mm}$ throughout the cell, tuned near the $2^3S_1 \rightarrow 2^3P_0$ transition.  The level structure of this transition is depicted in figure \ref{experiment_diagram}(b). In spin $1/2$ systems, spin fluctuations are associated to circular birefringence noise, and thus to stochastic Faraday rotation of the light polarization \cite{Zapasskii_review_2013, sinitsyn_theory_2016}. Spin-1 systems such as metastable Helium exhibit richer dynamics due to higher-order tensor spin degrees of freedom, resulting in extra spectral features such as resonances at twice the Larmor frequency \cite{Liu_2022, PhysRevA.107.023527}.

\begin{figure}
    \centering
    \includegraphics[width=0.9\columnwidth]{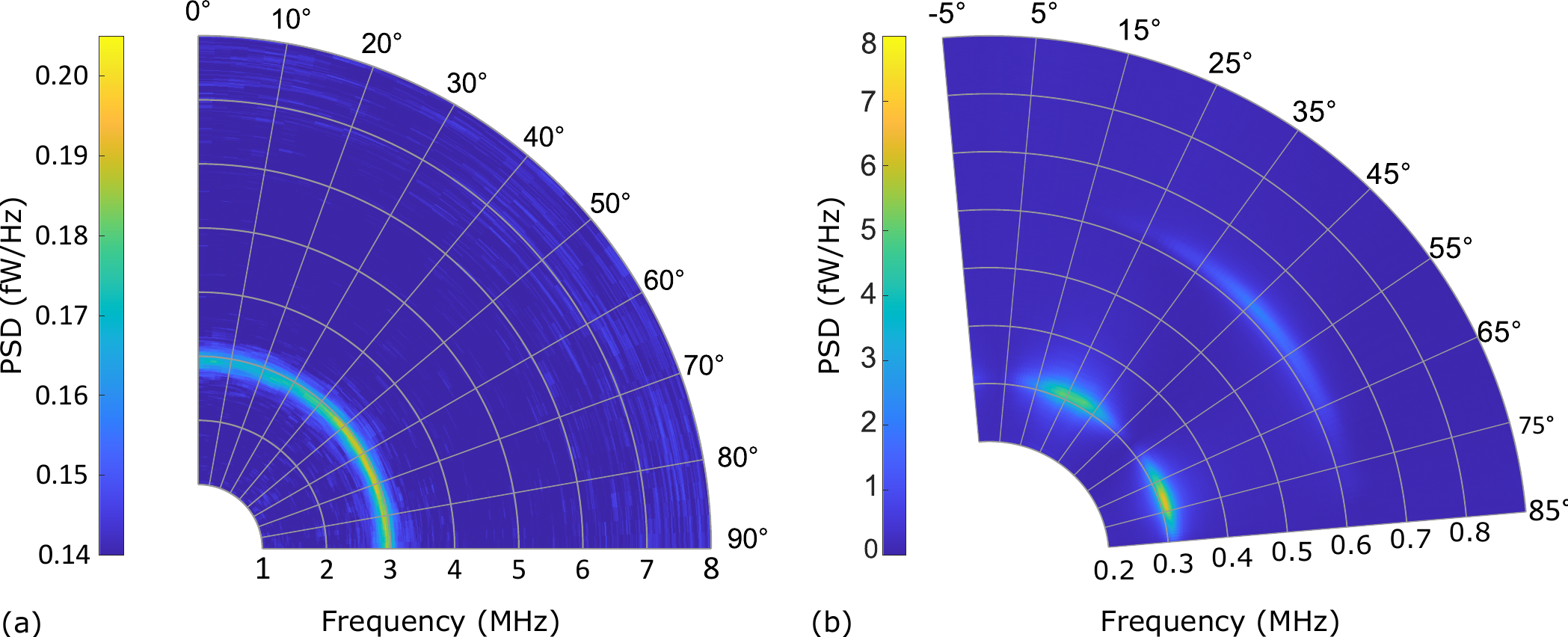}
    \caption{(a) Spin noise spectra measured in a well isolated environment. Experimental parameters are $P_{\mathrm{in}} = 1.5\, \mathrm{mW}$, $\Delta/2\pi = 1500\, \mathrm{MHz}$, $\omega_\mathrm{L}/2\pi = 3\,\mathrm{MHz}$. The probe polarization angle $\theta$ with respect to the magnetic field direction  is varied in the range $0$ to $90^{\circ}$. Such spectra are consistent with the theoretical expectation of a noise independent of $\theta$. (b) Spin noise spectra obtained in a environment where stray magnetic fields were likely to appear (prior to the lab shift).  Experimental parameters: probe power $P_{\mathrm{in}} = 3\, \mathrm{mW}$,  detuning $\Delta/2\pi = 300\, \mathrm{MHz}$. The Larmor frequency $\omega_\mathrm{L}/2\pi$ is around $0.3\, \mathrm{MHz}$. Above $200\, \mathrm{kHz}$, technical noises are suppressed and the flat background is dominated by the shot noise.}
    \label{PSD_experimentales}
\end{figure}

Figure \ref{PSD_experimentales}(a) shows a polar plot of recent FR noise measurements. Each radius displays the PSD corresponding to an angle $\theta$ between the probe polarization and the transverse magnetic field. This angle is varied between $0^{\circ}$ and $90^{\circ}$. One can see a resonance near the Larmor frequency $\omega_\mathrm{L}/2\pi \simeq 3\, \mathrm{MHz}$. The noise level is nearly the same whatever $\theta$, which is consistent with circular birefringence fluctuations. However, the results plotted in figure \ref{PSD_experimentales}(b), which were obtained when our laboratory was installed in an other building, are surprisingly different: no noise is observed when $\theta = 0,\, 50,$ or $90^{\circ}$, and the values reached for $\theta = 30,\, 70^{\circ}$ are also much higher. The larger probe power cannot explain such a difference of more than one order of magnitude.

A possible explanation for these striking differences is that the former building was much less isolated from surrounding stray magnetic fields. In this paper, we thus propose to assess the effect of a noise of the transverse magnetic field. We study the spin fluctuations that it can induce, hiding or spoiling the intrinsic, thermally induced spin noise. We wonder in the following section whether such fluctuations can successfully explain some of the features of figure \ref{PSD_experimentales}(b), and provide fundamental insight on the dynamics of spins in stochastic external fields.

\section{Theoretical model and simulations of SNS results}

We present in this section the  model we developed for the simulation of SNS experimental results. We focus here on the situation where the light probes spin noise near a $J=1 \rightarrow J=0$ transition, which  corresponds to the experiments presented in figure \ref{PSD_experimentales}. The general model for simulating the time evolution of the density matrix $\rho(t)$ of the open atomic system is detailed in Ref.\,\cite{Liu_2022}.  We focus in the following on the creation of a measurable spin noise signal due to the magnetic fluctuations solely, that is, without taking any other source of stochastic fluctuations of the density matrix elements into account, with noise levels comparable with the one observed experimentally where transit noise occurs.

\subsection{Theoretical model and numerical resolution scheme}

We investigate here the case of a uniaxial noisy magnetic field $\mathbf{B}(t) = (B+\delta B(t))\, \mathbf{e_x}$ oriented along the $x$ axis, where $\delta B(t)$ holds for the zero average fluctuations of the field amplitude, and  $\mathbf{e_x}$ is the unit vector along the $x$ direction. We assume in the rest of the paper that $\delta B(t) \ll B$. This results for the atoms in a fluctuating Larmor frequency $\omega_\mathrm{L}(t) = \gamma_1 B(t)=\omega_\mathrm{L} + \delta \omega_\mathrm{L}(t)$, where $\gamma_1 = g_1\mu_B/\hbar$ is the gyromagnetic ratio, and $g_1$ is the Land\'e factor of the lower level with $J=1$. We start from a Liouville-Von Neumann-like equation, including the Hamiltonian part of the evolution as well as the dissipation of the system: 
\begin{equation}
    \frac{\mathrm{d}\rho}{\mathrm{d}t}=\frac{1}{i\hbar}[H(t),\rho]+\frac{1}{i\hbar}{D}(\rho)\ .\label{eq1}
\end{equation}
The Hamiltonian H contains the coupling between lower Zeeman sublevels, through the time dependent Larmor frequency $\omega_{\mathrm{L}}(t)$, and the light-matter interaction between the probe and the atoms. The light, propagating in the $z$ direction, excites the $\sigma_{\pm}$ transitions with Rabi frequencies $\Omega_{\pm}$, and is detuned by $\Delta$ from the center of the transition  (see figure \ref{experiment_diagram}(b)). In the basis $\{ \ket{-1}_z, \ket{0}_z, \ket{+1}_z, \ket{e}\}$ where $\ket{i}_z$ is the Zeeman sublevel corresponding to a projection $m_z=i\hbar$ of the angular momentum along the $z$ axis and $\ket{e}$ is the excited level, $H$ reads:
\begin{equation}
 H=\hbar\left(\begin{array}{cccc}
 0 & \frac{\omega_\mathrm{L}(t)}{\sqrt{2}} & 0 & \frac{\Omega_+^{*}}{\sqrt{3}}\\
\frac{\omega_\mathrm{L}(t)}{\sqrt{2}} & 0 & \frac{\omega_\mathrm{L}(t)}{\sqrt{2}} & 0\\
 0 & \frac{\omega_\mathrm{L}(t)}{\sqrt{2}} & 0 & -\frac{\Omega_-^{*}}{\sqrt{3}}\\
\frac{\Omega_{+}}{\sqrt{3}} & 0 & -\frac{\Omega_{-}}{\sqrt{3}} & \Delta
\end{array}\right)\\ .\label{eq2}
\end{equation}

The dissipation matrix takes into account the spontaneous emission rate $\Gamma_0$, the dipole relaxation rate $\Gamma$, and the spin population and coherences relaxation rate $\gamma_t$ towards equilibrium, assuming it is dominated by the transit time of the atoms through the beam :
\begin{equation}
    D(\rho)= \\-i\hbar \left(\begin{array}{cccc}\gamma_t (\rho_{\text{-}1\text{-}1} -\frac{1}{3})-\frac{\Gamma_0}{3} \rho_{ee} & \gamma_t \rho_{\text{-}10} & \gamma_t \rho_{\text{-}11}& \Gamma \rho_{\text{-}1e} \\
    \gamma_t\rho_{0\text{-}1} & \gamma_t (\rho_{00} -\frac{1}{3})-\frac{\Gamma_0}{3} \rho_{ee} & \gamma_t\rho_{01} & \Gamma\rho_{0e} \\
    \gamma_t \rho_{1\text{-}1}& \gamma_t \rho_{10}& \gamma_t (\rho_{11} - \frac{1}{3})-\frac{\Gamma_0}{3} \rho_{ee} & \Gamma \rho_{\text{-}1e}\\
    \Gamma \rho_{e\text{-}1}& \Gamma \rho_{e0}& \Gamma \rho_{e1}& \Gamma_0 \rho_{ee}\\\end{array} \right)\ .
\label{eq3}
\end{equation}

We then write the density matrix $\rho$ under the 16-component vector form $\sigma = \left[ \rho_{\text{-}1\text{-}1} \, \rho_{\text{-}10} \,\rho_{\text{-}1+1} \,... \rho_{e\text{+}1}\,\rho_{ee}\right]^T$, so that (\ref{eq1}) can be cast in the following form:
\begin{equation}
\dfrac{\mathrm{d}\sigma}{\mathrm{d}t}=\left(\mathcal{\bar{L}} + \delta \mathcal{L}(t)\right) \sigma + \eta\ ,
\label{eq4}
\end{equation}
where $\mathcal{\bar{L}}$ contains the deterministic part $\bar{H}$ of $H(t)=\bar{H} + \delta H(t)$ as well as the dissipation terms. The term $\delta \mathcal{L}(t)$ corresponds to the fluctuating Hamiltonian $\delta H(t)$. The last term $\eta$ stands for the population feeding rate corresponding to the transit of the atoms through the beam. Indeed, moving atoms enters the laser volume with an average rate $\gamma_t$. Since they are in one of the three lower Zeeman sublevels with equal probabilities at thermal equilibrium, this leads to an increase in the lower Zeeman populations with a constant rate $\gamma_t/3$, while the exit of the atoms in included in the decay matrix $D(\rho)$. Moreover, because atoms enter the interaction volume in a well-defined sublevel of the lower state, they do not contribute to coherences and excited population terms. We thus write $\eta = [ \gamma_t/3, 0, ..., 0,\gamma_t/3, 0, ..., 0,\gamma_t/3,0,..., 0]$.

Contrary to simulations performed in Ref. \cite{Liu_2022}, we do not consider the fluctuations of the populations of the Zeeman sublevels, which are responsible for the standard spin noise. Since we focus on the effect of the fluctuating magnetic field, we just consider an average transit rate: the vector $\eta$ is constant, so that is does not by itself give rise to population imbalance. We then diagonalize $\mathcal{\bar{L}} = P\Lambda P^{-1}$ and make a change of variable $\Tilde{\sigma} = P^{-1} \sigma$, $\Tilde{\eta} = P^{-1}\eta$ so that, at first order of perturbation in $\delta \omega_\mathrm{L}(t)/\omega_\mathrm{L}$, we can write eq.\,(\ref{eq4}) as:
\begin{equation}
\dfrac{\mathrm{d}\Tilde{\sigma}}{\mathrm{d}t}=\left[\Lambda + \delta \omega_{\mathrm{L}}(t) \beta \right] \Tilde{\sigma} + \Tilde{\eta}\ ,
\label{Eq5}
\end{equation}
with $\beta$ a diagonal matrix corresponding to the derivative of $\Lambda$ with respect to $\omega_L$:
\begin{equation}
    \beta = \dfrac{\Lambda (\omega_{\mathrm{L}}+\mathrm{d}\omega_{\mathrm{L}}) - \Lambda (\omega_{\mathrm{L}}-\mathrm{d}\omega_{\mathrm{L}}) }{2\,\mathrm{d}\omega_{\mathrm{L}}}\ ,
    \label{eq6}
\end{equation} 
and $\delta\omega_{\mathrm{L}}(t)$ a numerically generated Gaussian random noise term. From our definition of H and $D(\rho)$, the numerical computations show that $\Lambda$ has imaginary coefficients with negative real parts, and $\beta$ shows coefficients with non-negligible imaginary part only.

Eventually, the system is integrated:
\begin{equation}
\Tilde{\sigma}(t) = \Tilde{\sigma}(0){\mathrm{e}}^{\Lambda t+\beta \int_0^t\delta \omega_{\mathrm{L}}(t')\,\mathrm{d}t'} +\int_0^t \mathrm{d}t'\, \Tilde{\eta}\, {\mathrm{e}}^{\Lambda(t-t')} {\mathrm{e}}^{\beta \int_{t'}^t \delta \omega_{\mathrm{L}}(t'')\, \mathrm{d}t''}\ .
\label{eq7}
\end{equation}

The density matrix $\sigma = P\Tilde{\sigma}$ and thus the atomic polarization of the system are computed at each time, and the spin noise spectra are then simulated according to the method presented in Ref.\,\cite{Liu_2022}. 

Interestingly, we can already conclude from (\ref{eq7}) that no spin noise can be created by the magnetic field fluctuations if the system is closed, i.e. if there is no transit of the atoms through the beam. Indeed, in this case, the second term in (\ref{eq7}) vanishes, and the random phase of the first term  averages out after a sufficiently long time of interaction, since $t^{-1} \int_0^t\delta \omega_{\mathrm{L}}(t')\,\mathrm{d}t' \rightarrow 0$ when $t \rightarrow \infty$ for a centered random process. On the contrary, atoms flying in and out of the interaction volume interact with the beam within a finite time, and the acquired random phase does not average down to 0. Such a physical picture is depicted by the second term in eq.(\ref{eq7}), where atoms entering the beam at time $t'$ can acquire a non-negligible random phase $\int_{t'}^t \delta \omega_{\mathrm{L}}(t'')\, dt''$ at time $t$, allowing a stationary noise to exist.

\subsection{Emergence of Faraday rotation noise: fluctuations of $S_z(t)$}

We focus on the case of a stationary, Gaussian, and correlated magnetic noise, thus fully characterized by its autocorrelation function \cite{kampen_stochastic_2007}

\begin{equation} \overline{\delta\omega_\mathrm{L}(t') \delta\omega_\mathrm{L}(t)} =  \omega_\sigma^2 \exp \left(  - \vert t'-t\vert/\tau_{\mathrm{c}}\right)\ ,  \label{eq8}
\end{equation}
with $\omega_\sigma$ the standard deviation of the Larmor frequency noise and $\tau_{\mathrm{c}}$ its correlation time. The top bar here denotes statistical ensemble average. Numerically, we first simulate a Gaussian noise with a correlation time equal to the time step $\mathrm{d}t$ of our simulation scheme. Since  $1/\mathrm{d}t$ exceeds all other typical frequencies of the process, this creates an approximately white noise. We then apply a frequency filter with a Lorentzian shape, centered around zero frequency and with a half width at half maximum (HWHM) given by $1/\tau_{\mathrm{c}}$.

\begin{figure}[ht]
    \centering
    \includegraphics[width=0.45\columnwidth]{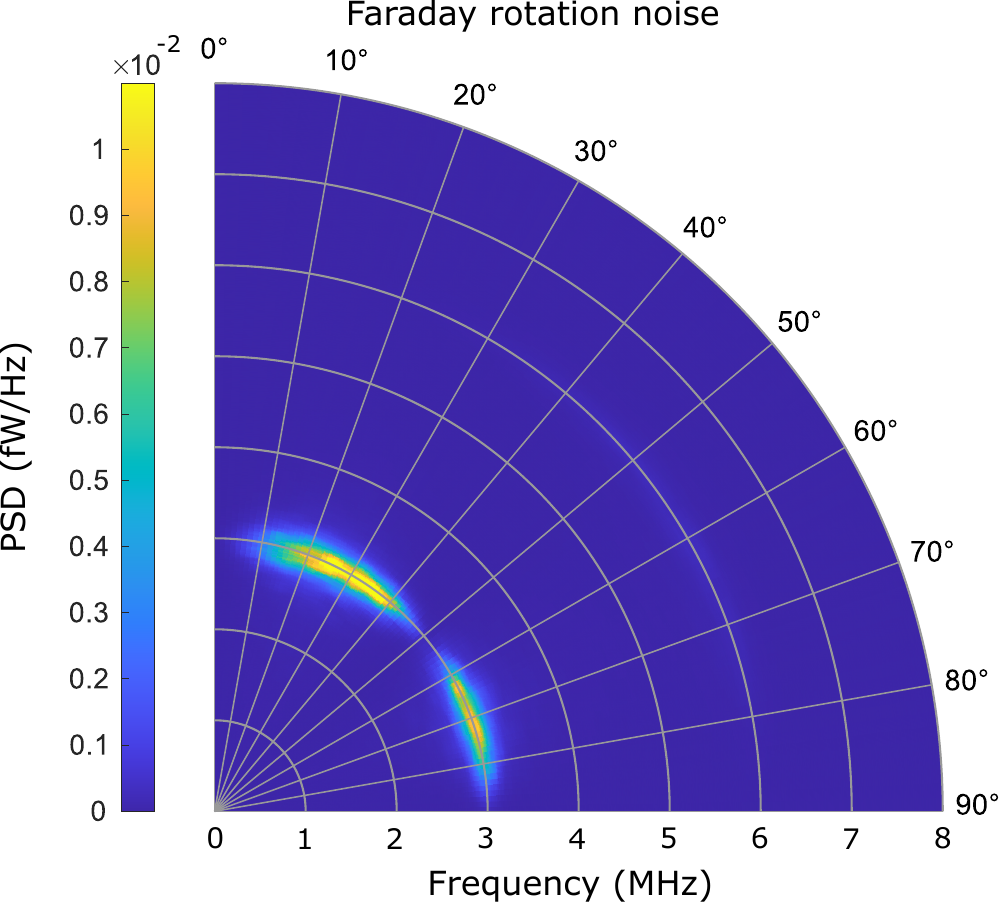}
    \caption{Rotation noise spectra obtained for $P = 1.5\,\mathrm{mW}$, $\Delta /2\pi = 1500\,\mathrm{MHz}$, $\omega_{\mathrm{L}} / 2\pi=3\,\mathrm{MHz}$, and for $\theta$ varying from 0 to 90$^{\circ}$. The magnetic noise correlation time is $\tau_{\mathrm{c}} = 5.3\times 10^{-9}\,\mathrm{s}$, its standard deviation corresponds to 12\% of $\omega_{\mathrm{L}}$.}
    \label{simu_rot}
\end{figure}

The results of the simulated Faraday rotation spectra are represented in figure \ref{simu_rot} in polar coordinates, with an angle $\theta$ between the light polarization and the magnetic field between 0 and $90^{\circ}$. The values of the parameters are extracted from previous experimental works: $\Gamma_0/2\pi= 1.63\,\mathrm{MHz}$, $\Gamma / 2\pi=  800\,\mathrm{MHz}$, $\gamma_t /2\pi= 60\,\mathrm{kHz}$. The probe beam optical detuning and Rabi frequency are $\Delta/2\pi = 1500\,\mathrm{MHz}$ and $\Omega/ 2\pi = 50\,\mathrm{MHz}$ (obtained from a $1.5\,\mathrm{mW}$ laser power and a $0.6\,\mathrm{mm}$ beam diameter). The Gaussian magnetic noise has a standard deviation  $\omega_{\sigma} = 0.12\times \omega_{\mathrm{L}}$, and the correlation time of the noise is $\tau_{\mathrm{c}} = 5.3\times 10^{-9}\,\mathrm{s}$, corresponding to a bandwidth of $30 \, \mathrm{MHz}$. With these parameters, some non-negligible spin noise is efficiently created around the Larmor frequency, with a PSD comparable to the one obtained numerically and experimentally with transit noise (see Refs. \cite{Liu_2022, PhysRevA.107.023527, PhysRevA.107.L011701} for such results). Moreover, no noise is visible for $\theta = 0, \,50^{\circ}, \, 90^{\circ}$: figure \ref{simu_rot} is thus very similar to figure \ref{PSD_experimentales}(b). This behavior is different from the isotropy of standard circular birefringence noise (visible on the contrary on figure \ref{PSD_experimentales}(a) and reported in \cite{PhysRevA.107.023527}): this effect will be discussed in section 3.

\begin{figure}[h]
    \centering
    \includegraphics[width = 0.50\columnwidth]{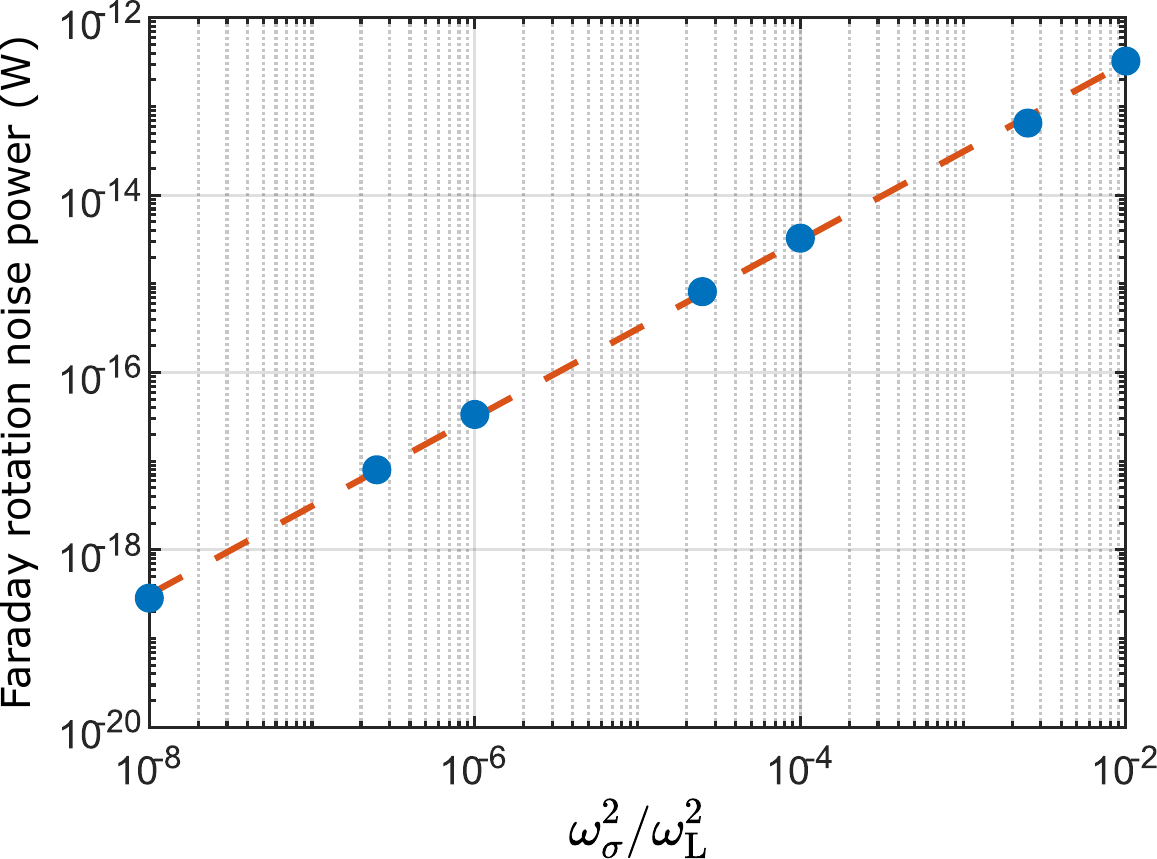}
    \caption{Variance of the Faraday rotation noise, obtained for $P = 1.5\,\mathrm{mW}$, $\Delta/2\pi = 1500\,\mathrm{MHz}$, and $\theta = 30^{\circ}$. The magnetic noise correlation time is $\tau_{\mathrm{c}} = 5.3\times10^{-9}\,\mathrm{s}.$ Blue dots : simulations; orange dashed line: power law fit $y=ax^k$. The exponent is found to be $k=0.9987$, very close to 1.}
    \label{Variance_vs_PSDmag}
\end{figure}

To assess the impact of the power of the magnetic noise, we simulate similar Faraday rotation noise spectra (for a fixed angle $\theta = 30 ^{\circ}$) with a variance of the Larmor frequency noise $\omega_\sigma^2$ varying very broadly from  of $10^{-8}\times \omega_{\mathrm{L}}^2$ to $10^{-2}\times \omega_{\mathrm{L}}^2$. The results can be seen on figure \ref{Variance_vs_PSDmag}. The integrated PSD (i.e. the variance of the simulated FR noise) is plotted on a log-log scale, as a function of the ratio $\omega_\sigma^2/\omega_{\mathrm{L}}^2$ (blue dots). The data are fitted by a power law function $y=ax^k$ (orange dash-dotted line), whose exponent is found to be $k=1$, thus proving that the spin noise variance is proportional to the magnetic noise variance $\omega_\sigma^2$.

These numerical results give an idea of the levels of additional spin noise that can accidentally be created by a stray magnetic field. However, they give little physical insight into the mechanisms from which this spin noise originates. In particular, the fact that it can emerge from noise in the Hamiltonian only, and not from stochastic fluctuations of the density matrix elements, is surprising and quite unclear at this stage. In the following, we develop a theoretical model that aims at providing a clear physical interpretation to this observation. We also explain the specific features highlighted in this section, like the dependence on the statistical properties of the magnetic noise and the specific polarization dependence of the simulated spectra.

\section{Analytical solution : perturbative treatment and physical discussion }
To provide further insight into the numerical simulation results, we derive in this section analytical expressions for the spin correlator and the variance of the spin noise, that can be interpreted physically to explain the mechanisms of creation of these fluctuations. We thus clarify the conditions for the appearance of such a spin noise with respect to the central frequency of the magnetic fluctuations and their bandwidth. We finally attempt to shed some light on the role of the residual optical pumping in the existence of this noise.

\subsection{Spin equation of motion}

We start again from the master equation  (\ref{eq1}), in which we write $D(\rho) = D'(\rho) + \eta$, with $\eta$ containing the average feeding terms due to the transit, as we did in eq.(\ref{eq4}) :

\begin{equation}
    \frac{\mathrm{d}\rho}{\mathrm{d}t}=\frac{1}{i\hbar}[H(t),\rho]+\frac{1}{i\hbar}{D'}(\rho) + \frac{1}{i\hbar}\eta\, . \label{eq9}
\end{equation}
We then apply a perturbative treatment: under the action of the deterministic magnetic field only and the relaxation processes included in $D(\rho)$, the density matrix of the system reaches a steady state $\rho^{\mathrm{st}}$, solution of $ \left[ \bar{H}, \rho^{\mathrm{st}}\right] +D(\rho^{\mathrm{st}}) + \eta =0$. We study the fluctuations of the density matrix around steady-state at first order in $\delta \omega_\mathrm{L}(t) \ll \omega_\mathrm{L}$. We write $\rho(t) = \rho^{\mathrm{st}} + \delta\rho$, with $\delta\rho$ the solution of the master equation developed at first order:

\begin{equation}
    \frac{\mathrm{d} \delta\rho}{\mathrm{d}t} -\frac{1}{i\hbar}[\bar{H},\delta\rho] -\frac{1}{i\hbar}{D'}(\delta\rho) = \frac{1}{i\hbar}[\delta H(t),\rho^{\mathrm{st}}]\,.
    \label{eq10}
\end{equation}
This equation can be interpreted in the following way : the perturbation $\delta \rho(t)$ sees its dynamics ruled by the deterministic hamiltonian $\bar{H}$ and relaxation $D'(\delta\rho)$ (without feeding terms, which are of order 0), and undergoes stochastic forcing terms induced by the fluctating part of the Hamiltonian acting on the steady-state $\rho^{\mathrm{st}}$. Under the approximation of a relaxation rate $\gamma$ for the the populations and coherences due to the transit rate of the atoms, one can integrate equation (\ref{eq10}). The calculation is detailed in Appendix A. This yields: 
\begin{equation}
\delta \rho(t) = -\dfrac{i}{\hbar}\int_{-\infty}^t\mathrm{d}t' \,\left[ \delta H (t'),\, {\mathrm{e}}^{-i\bar{H}(t-t')/\hbar}\,\rho^{\mathrm{st}}\, {\mathrm{e}}^{i\bar{H}(t-t')/\hbar}\right] e^{-\gamma(t-t')}\ .
    \label{eq11}
\end{equation}

\subsection{Emergence of Faraday rotation noise : fluctuations of $S_z(t)$}

To investigate the spin noise created by such a motion, we operate the expansion of the density matrix over the eight degrees of freedom corresponding to the matrices $M_i$ that to some extent generalize Pauli matrices introduced in Refs. \cite{colangelo_quantum_2013} and \cite{PhysRevA.107.023527}. These matrices are generators of the SU(3) group, and up to a change of basis, correspond to the Gell-Mann matrices used in particle physics \cite{georgi2000lie}.  We thus write:
\begin{equation}
\rho^{\mathrm{st}} = \dfrac{1}{3} \mathbf{1} + \dfrac{1}{2} \sum_{i=1}^8 \lambda_i^{(\mathrm{st})} M_i
    \label{eq12}
\end{equation}
and \begin{equation}
\delta\rho =  \dfrac{1}{2} \sum_{i=1}^8 \lambda_i^{(1)} M_i
    \label{eq13}
\end{equation}

We remind that the $M_i$'s are traceless, orthogonal, and Hermitian operators, and that $\lambda_i = \mathrm{Tr}(\rho M_i) = \langle M_i\rangle$. More details, as well as the explicit forms of those matrices, can be found in Appendix B. The three first matrices correspond to the spin operators, which we denote $M_1 = S_z$, $M_2 = S_x$, and $M_3 = S_y$. The remaining operators $M_{4...8}$ account for tensorial degrees of freedom, and are the subject of section 4. Here, we focus on the Faraday rotation noise, which emerges from fluctuation in $S_z$. The deterministic Hamiltonian writes $\bar{H} = \hbar \omega_\mathrm{L} M_2$. Using the decomposition of eq.\,(\ref{eq13}) we write again eq.\,(\ref{eq11}) as:

\begin{equation}
    \delta \rho(t) = -\dfrac{i}{2} \sum_{j=1}^8 \lambda_j^{(\mathrm{st})}\int_{-\infty}^t\mathrm{\mathrm{d}t'} \delta\omega_\mathrm{L}(t')\,\left[ M_2,\, {\mathrm{e}}^{-iM_2\omega_\mathrm{L}(t-t')}\,M_j\, {\mathrm{e}}^{iM_2\omega_\mathrm{L}(t-t')}\right] \exp^{-\gamma(t-t')}\ .
    \label{eq14}
\end{equation}

The commutator $\left[ M_2,\, {\mathrm{e}}^{-iM_2\omega_\mathrm{L}(t-t')}\,M_j\, {\mathrm{e}}^{iM_2\omega_\mathrm{L}(t-t')}\right]$ maps the operators $M_1 = S_z$ and $M_3 = S_y$ on each other: this physically corresponds to the precession of the spin in the $(y,z)$ plane under the action of the magnetic field. Thus, the equations of motion for $\lambda_z(t)$ and $\lambda_y(t)$ fully characterize the stochastic evolution of $\lambda_z(t)$. They are coupled by the $\mathbf{B}$ field according to : 
\begin{equation}
    \left \{
    \begin{array}{c c}
     \lambda_z^{(1)}(t) =& \lambda_y^{(\mathrm{st})} \int_{-\infty}^t \mathrm{dt'}\delta\omega_\mathrm{L}(t') \cos \omega_\mathrm{L} (t-t') {\mathrm{e}}^{-\gamma (t-t')} \\
        &-\lambda_z^{(\mathrm{st})} \int_{-\infty}^t \mathrm{dt'} \delta\omega_\mathrm{L}(t') \sin \omega_\mathrm{L} (t-t') {\mathrm{e}}^{-\gamma (t-t')}\ , \\
      & \\
     \lambda_y^{(1)}(t) =& -\lambda_z^{(\mathrm{st})} \int_{-\infty}^t \mathrm{dt'}\delta\omega_\mathrm{L}(t') \cos \omega_\mathrm{L} (t-t') {\mathrm{e}}^{-\gamma (t-t')} \\
     &-\lambda_y^{(\mathrm{st})} \int_{-\infty}^t \mathrm{dt'} \delta\omega_\mathrm{L}(t') \sin \omega_\mathrm{L} (t-t') {\mathrm{e}}^{-\gamma (t-t')} \ .
    \end{array} 
    \right.
    \label{eq15}
\end{equation}

Such explicit equations can be integrated to get the expression of the correlator corresponding to the FR noise, i. e., $\overline{\lambda_z^{(1)}(T) \lambda_z^{(1)}(0)}$. We stick to the case of a Gaussian, correlated magnetic noise with variance $\omega_\sigma^2$ and bandwidth $1/\tau_{\mathrm{c}}$. In the regime where the relaxation rate $\gamma$ is smaller than the Larmor frequency and the bandwidth of the magnetic noise, one obtains (the calculations are detailed in appendix C):
\begin{equation}
    \overline{\lambda_z^{(1)}(T) \lambda_z^{(1)}(0)} = \lambda_+^{(\mathrm{st})}\lambda_-^{(\mathrm{\mathrm{st}})} \dfrac{\omega_\sigma^2\tau_{\mathrm{c}}}{2\gamma} \dfrac{1}{1+\omega_{\mathrm{L}}^2\tau_{\mathrm{c}}^2} \cos(\omega_{\mathrm{L}} T){\mathrm{e}}^{-\gamma\vert T\vert}\ .
    \label{eq16}
\end{equation}

\begin{figure}
    \centering
    \includegraphics[width=0.55\columnwidth]{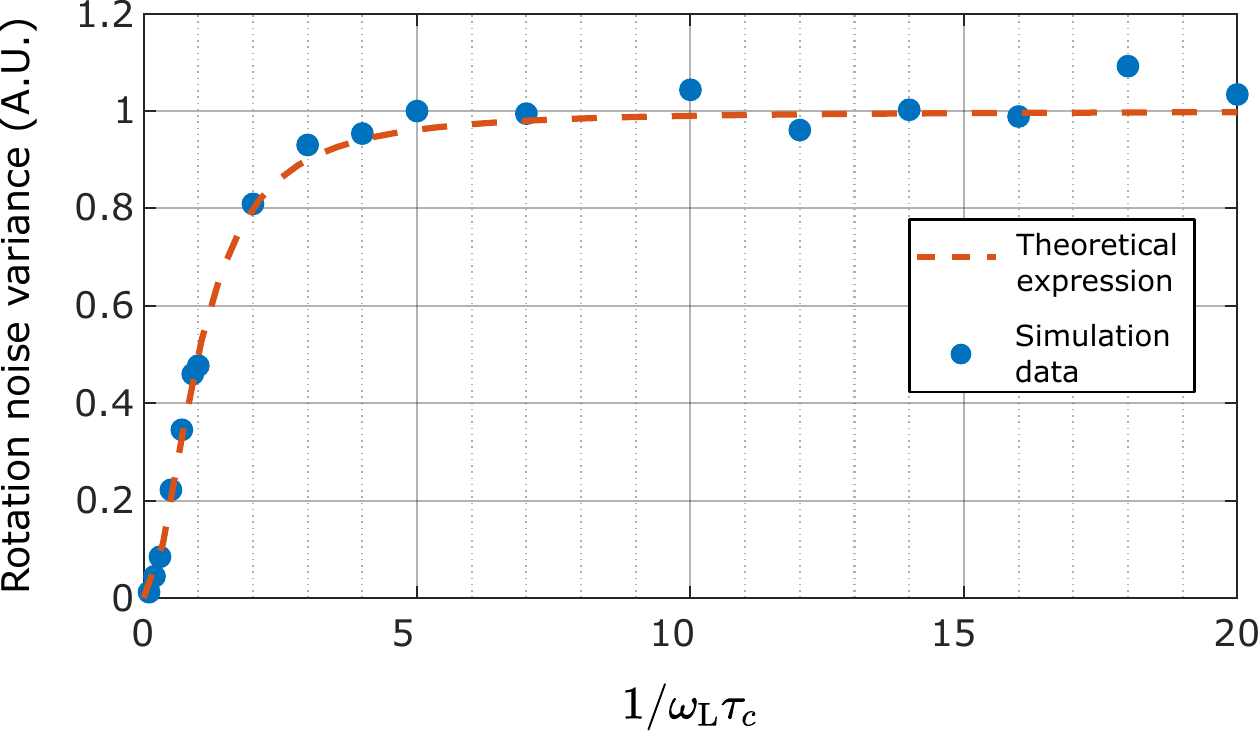}
    \caption{Variance of the Faraday rotation noise induced by the  spin fluctuations $S_z(t)$ as a function of the magnetic noise bandwidth $1/\tau_{\mathrm{c}}$ (in units of $\omega_\mathrm{L}$). The product $\omega^2_\sigma \tau_{\mathrm{c}}$ is kept constant for all points. The parameters are : $P = 1.5\,\mathrm{mW}$, $\Delta/2\pi = 1500\,\mathrm{MHz}$, $\theta = 30^{\circ}$. Blue dots: simulations; orange dashed line: theoretical expression derived from (\ref{eq16}).}
    \label{unmodulated_noise}
\end{figure}

We directly see from (\ref{eq16}) that the variance of the spin noise is given by $\mathrm{Var} \,S_z = \lambda_+^{(\mathrm{st})}\lambda_-^{(\mathrm{st})} \dfrac{\omega_\sigma^2\tau_{\mathrm{c}}}{2\gamma} \dfrac{1}{1+\omega_{\mathrm{L}}^2\tau_{\mathrm{c}}^2}$. We thus prove theoretically the proportionality between the spin noise and the magnetic noise variances, as observed numerically in figure \ref{Variance_vs_PSDmag}. On the other hand, the product $\omega_\sigma^2\tau_{\mathrm{c}}$ corresponds to the average power spectral density of the  magnetic noise. By varying the correlation time $\tau_{\mathrm{c}}$ while keeping this product constant, one can investigate the role of the noise bandwidth while keeping the maximum value of the magnetic field PSD constant. In this case, one can see from (\ref{eq16}) that the spin noise variance scales like $f(x) = 1/\left[1+(1/x)^2\right]$ with $x=1/\omega_\mathrm{L}\tau_{\mathrm{c}}$ (i.e. the bandwidth of the magnetic noise expressed in dc Larmor frequency units). This behaviour is shown in figure \ref{unmodulated_noise}, which reproduces the simulated  spin noise variance as a function of an increasing bandwidth of the magnetic noise, in the range $0.3\omega_\mathrm{L}$ to $20\omega_\mathrm{L}$ (blue dots). One can see that for very narrow bandwidths, no spin noise is excited. However, as soon as $1/\tau_{\mathrm{c}}$ becomes of the order of $\omega_\mathrm{L}$, the magnetic noise power in the frequency band containing $\omega_\mathrm{L}$ starts increasing. It then contributes quadratically to the Faraday rotation noise. This FR noise eventually saturates and reaches its maximum variance when $1/\tau_{\mathrm{c}} \simeq 10\,\omega_\mathrm{L}$. The data are normalized by this maximum variance, which indeed does not depend on $\tau_{\mathrm{c}}$. For comparison, the orange dashed line represents the function $f(x) = 1/\left[1+(1/x)^2\right]$, derived from our analytical expression eq.\,(\ref{eq16}). The agreement with the simulation is excellent, thus validating our first order perturbation model.

\subsection{Case of a noise modulated at a frequency $\Omega$}
The influence of the magnetic noise bandwidth can be understood physically: equation (\ref{eq11}) is similar to that of a set of coupled driven oscillators corresponding to the different elements of the density matrix. The forcing terms $[\delta H(t),\Tilde{\rho}^{\mathrm{st}}]/i\hbar$ contain a broad band of frequency components, going from 0 to $1/\tau_{\mathrm{c}}$. However, only the power density in the band containing the natural oscillator frequency, i.e. the dc Larmor frequency $\omega_\mathrm{L}$ can resonantly excite the spin noise. Therefore, the frequency components of the magnetic fluctuations that excite spin noise are not determined by an intrinsic, atomic parameter, but rather by tunable experimental conditions. This explains why no noise is visible for narrow bandwidths, and why it saturates when the bandwidth gets much broader than $\omega_\mathrm{L}$  while keeping the PSD constant. To emphasize this point, one can also study the case of a modulated magnetic noise, which has its PSD centered at a finite frequency $\Omega$. The autocorrelation function of $\delta\omega_\mathrm{L}(t)$ becomes 
\begin{equation}
\overline{\delta \omega_\mathrm{L} (t') \delta\omega_\mathrm{L} (t)} = \omega_\sigma^2 \cos \Omega (t-t') \exp \left( \vert t'-t\vert/\tau_{\mathrm{c}}\right)\ ,
    \label{eq17}
\end{equation}
and one can show that the variance of the spin noise now reads
\begin{equation}
\mathrm{Var} \,S_z = \lambda_+^{(\mathrm{st})} \lambda_-^{(\mathrm{st})} \dfrac{\omega_\sigma^2\tau_{\mathrm{c}}}{2\gamma} \dfrac{1}{1+(\omega_\mathrm{L}-\Omega)^2\tau_{\mathrm{c}}^2}\ .
    \label{eq18}
\end{equation}

\begin{figure}
    \centering
    \includegraphics[width=0.55\columnwidth]{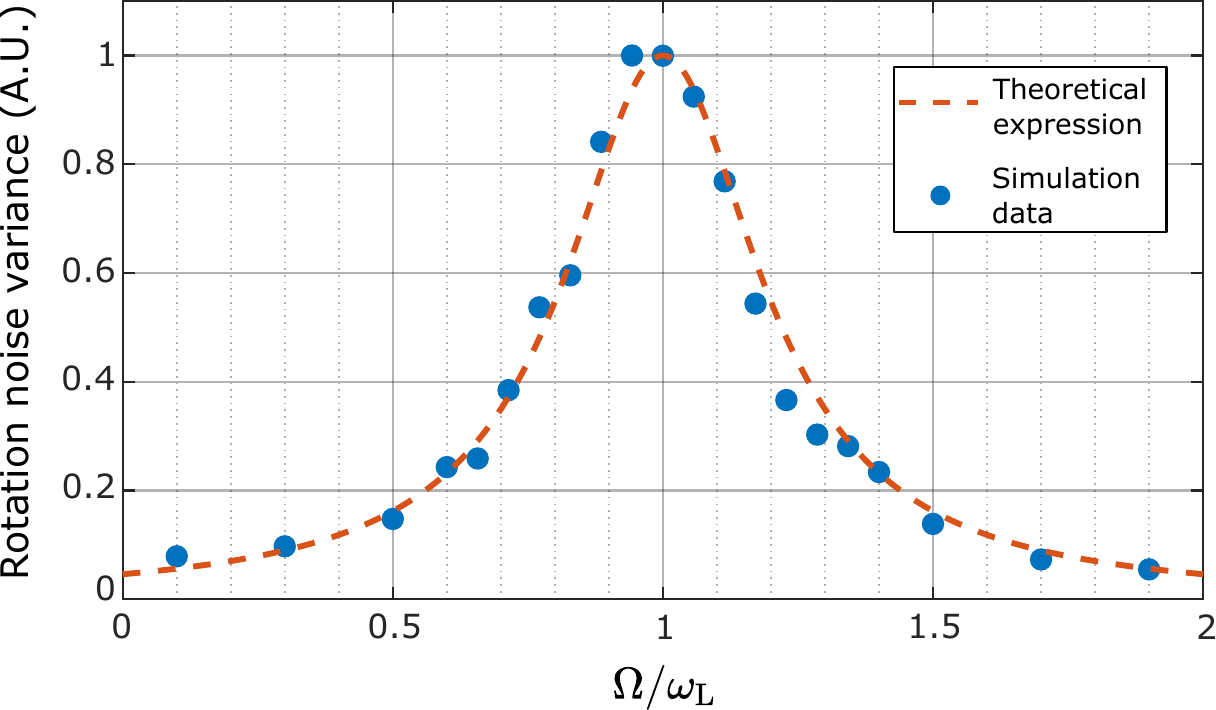}
    \caption{Variance of the Faraday rotation noise induced by the  spin fluctuations $S_z(t)$ as a function of the modulating frequency $\Omega$ in units of $\omega_\mathrm{L}$. The parameters are: $P = 1.5\,\mathrm{mW}$, $\Delta/2\pi = 1500\,\mathrm{MHz}$, $1/2\pi\tau_{\mathrm{c}} = 0.6\,\mathrm{MHz}$, $\theta = 30^{\circ}$. Blue dots: simulations ; orange dashed line: theoretical expression derived from (\ref{eq18}).}
    \label{modulated_noise}
\end{figure}

Thus, by fixing again the product $\omega_\sigma^2\tau_{\mathrm{c}}$, the variance of $S_z$ follows a Lorentzian evolution as a function of the modulation frequency $\Omega$: almost no spin noise is created if $\Omega \gg \omega_\mathrm{L}$ or $ \Omega \ll \omega_\mathrm{L}$, whatever the magnetic noise power. Figure \ref{modulated_noise} indeed shows the evolution of the simulated spin noise  variance (in blue dots) when the modulation frequency $\Omega$ is swept in the range $0-2\omega_\mathrm{L}$. As expected, some spin noise is efficiently created only if $\Omega \simeq \omega_\mathrm{L}$, with a resonance HWHM given by the bandwidth $1/\tau_{\mathrm{c}}$. The data are normalized to the maximum value, and the orange dashed line represents the normalized function $f(x) = 1/\left[1+(\omega_\mathrm{L}-\Omega)^2\tau_{\mathrm{c}}^2\right]$, derived from equation (\ref{eq18}). The agreement with the simulations is again excellent. This proves that only the power of the magnetic noise lying in the band around $\omega_\mathrm{L}$ creates the noise. Indeed, in the case of a zero-frequency centered noise, a noise with a level comparable with the one induced by transit could be obtained for a bandwidth of $30 \,\mathrm{MHz}$ and a standard deviation $\omega_\mathrm{\sigma}$ of 12\% of the dc B field. However, with a modulation at $\Omega = \omega_\mathrm{L}$, the same level can be obtained with both a much smaller bandwidth of $1/2\pi\tau_{\mathrm{c}} =  600  \,\mathrm{kHz}$ and a much smaller standard deviation of 2\% of $\omega_\mathrm{L}$. With a bandwidth of $ 100 \,\mathrm{kHz}$, this standard deviation drops to only 1\%. This naturally follows from the fact that the critical parameter is the power density of the magnetic noise at the Larmor frequency.
This case of a modulated noise thus shows that the induced spin noise can be measurable even with very low magnetic noise variance, as soon as the dc Larmor frequency matches the one of a nearby source of magnetic noise.

This model is consequently very helpful for interpreting the creation of spin noise in terms of random coupling induced by the magnetic noise between the spin degrees of freedom $M_j$ near steady-state. In the case of the Faraday rotation noise, this is equivalent to a stochastic precession of the spin in the $(y,z)$ plane around its steady-state orientation because of the noisy magnetic field. Equation (\ref{eq15}) shows that the steady-state has a strong impact, since in our perturbative model this coupling is always between an operator at first order and an other operator at steady-state. In the following, we discuss the origin of the steady-state in terms of optical pumping and its consequence on the polarization dependence of the noise.

\subsection{Influence of the optically pumped steady-state : polarization dependence of the Faraday rotation noise}

We now discuss the influence of the factor $\lambda_+^{(\mathrm{st})}\lambda_-^{(\mathrm{st})}$ in eq.\,(\ref{eq18}). First we rewrite $\lambda_+^{(\mathrm{st})}\lambda_-^{(\mathrm{st})}$ as 

\begin{equation}
    \lambda_+^{(\mathrm{st})}\lambda_-^{(\mathrm{st})} = (\lambda_z^{(\mathrm{st})})^2 + (\lambda_y^{(\mathrm{st})})^2 = \langle S_z ^{(\mathrm{st})}\rangle ^2 + \langle S_y ^{(\mathrm{st})}\rangle ^2 = \langle S_\perp^{(\mathrm{st})} \rangle ^2\ ,
    \label{eq19}
\end{equation}
where $\langle S_\perp^{(\mathrm{st})} \rangle ^2$ denotes the squared spin component in the plane $yz$, orthogonal to the magnetic field. Thus, (\ref{eq16}) shows that no spin noise is measurable if the steady-state spin of the system has no component in the transverse plane. This can be understood by realizing that there is no stochastic precession induced by the magnetic noise if the spin is aligned with the $\mathbf{B}$ field in the first place. 

\begin{figure}
\centering
    \includegraphics[width =0.50 \columnwidth]{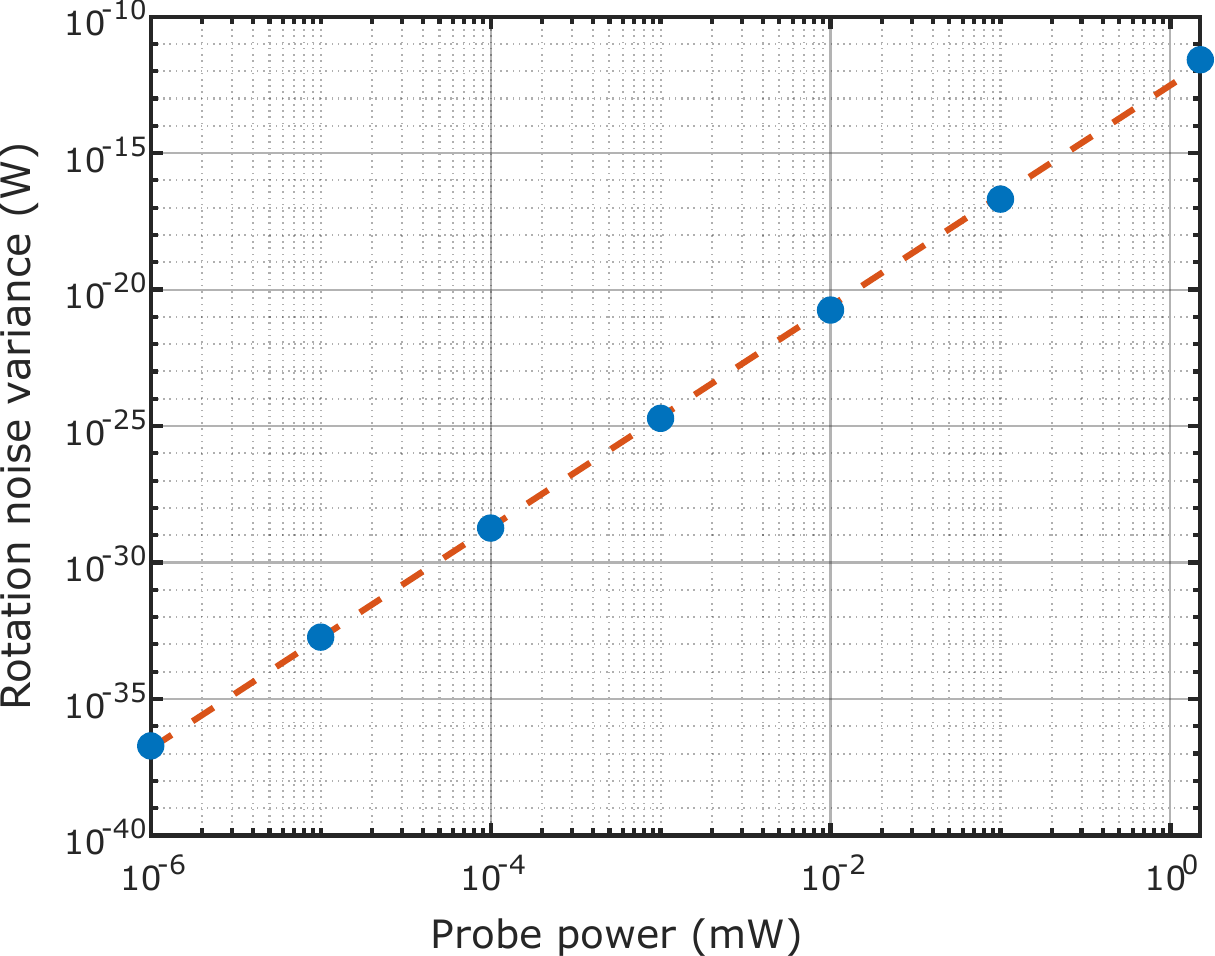}
    \caption{Variance of the  Faraday rotation noise for $\Delta/2\pi = 1500\,\mathrm{MHz}$ and $\theta = 30^{\circ}$, as a function of the probe power $P$. The magnetic noise correlation time is $\tau_{\mathrm{c}} = 5.3\times10^{-9}\,\mathrm{s}.$ Blue dots : simulations; orange dashed line: power law fit $y=ax^k$. The exponent is found to be $k=4.0018$, very close to 4.}
    \label{Variance_vs_ProbePower}
\end{figure}

Lets now discuss the origin of this transverse spin at steady-state. In most cases, one probes SNS in the wings of the absorption profile, which can result in a small absorption and de-excitation of the medium \cite{fomin_spin-alignment_2020, fomin_anomalous_2021}, leading to an activation of the degree of freedom $M_j$ at steady-state. We thus investigate the dependence of the spin noise variance on the laser probe power, at a fixed detuning $\Delta / 2\pi= 1500\, \mathrm{MHz}$ and a fixed angle $\theta = 30^{\circ}$ between the light polarization and the magnetic field. The results are shown in figure \ref{Variance_vs_ProbePower}: the blue dots are the simulation results and the orange dashed line is a power-law fit function $y(x)=ax^k$. While the standard, motion-induced spin noise signal in atomic vapors is well known to increase quadratically with the probe power $P$ \cite{bai_enhancement_2022, Mihaila_Crooker_Rickel_Blagoev_Littlewood_Smith_2006, glazov_linear_2015}, the simulated variance of the spin noise induced by the magnetic noise scales like $P^4$. This is perfectly consistent with an explanation in terms of residual absorption by the medium, leading to a small optical pumping: indeed, the probe laser then contributes both to the creation and to the detection of the spin noise.

\begin{figure}
    \centering
    \includegraphics[width=0.8\columnwidth]{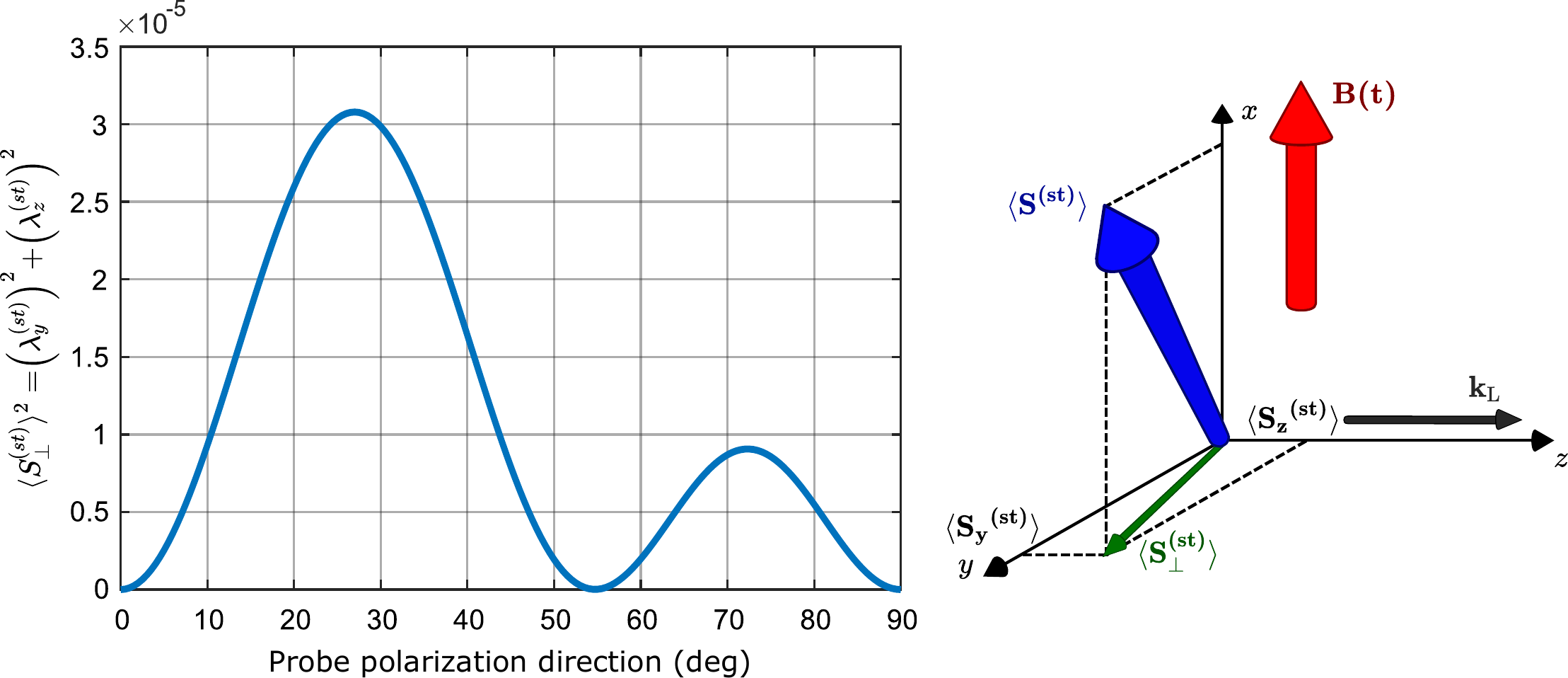}
    \caption{ Left: Evolution versus $\theta$ of the squared transverse spin component $\left(\lambda_y^{(\mathrm{st})}\right)^2 + \left(\lambda_z^{(\mathrm{st})}\right)^2 = \langle S_\perp^{(\mathrm{st})}\rangle^2$ obtained at steady-state for $P = 1.5\,\mathrm{mW}$, $\Delta/2\pi = 1500\,\mathrm{MHz}$, and $\omega_\mathrm{L}/2\pi= 3\,\mathrm{MHz}$. Right: schematic of the average spin of the sample (blue arrow) after the action of the probe light : residual optical pumping results in a spin having a non-zero component in the ($x$,$y$) plane (green arrow). $\mathbf{k_\mathrm{L}}$ stands for the probe laser wave vector, aligned along the z axis. }
    \label{stationary_transverse_spin}
\end{figure}

The proportionality between the spin noise power and $\langle S_\perp^{(\mathrm{st})} \rangle ^2$ also explains the very specific polarization dependence of the spin noise signals reproduced in figure \ref{simu_rot}. Indeed, in our model of a $J=1 \rightarrow J=0$ transition, the strength of the optical pumping  and the nature of the steady-state after a few cycles of pumping strongly depends on the directions of the laser polarization and of the magnetic field (see \cite{Liu_2022} for details). For instance, figure  \ref{stationary_transverse_spin} shows the squared transverse spin component at steady-state as a function of the angle $\theta$. The transverse component is maximum for $\theta \simeq 30 ^{\circ}$ and $70 ^{\circ}$. On the contrary, when the probe light is aligned with, orthogonal to, or at $55^{\circ}$ with respect to the $\mathbf{B}$ field, no transverse spin is created. One can then compare this evolution with figures \ref{simu_rot} and \ref{PSD_experimentales}(b): the angle for which no transverse optical pumping occurs matches perfectly the polarization directions where no noise is obtained. We recall here that the detection of the FR noise is isotropic: the absence of observable noise for certain values of $\theta$ unambiguously proves that no noise is created in these directions of polarization, since otherwise it would be necessarily observed.

We have thus successfully explained and interpreted the creation of the spin fluctuations that result in Faraday rotation noise, by providing analytical expressions based on the one hand on a perturbative treatment of the quantum equation of motion, and on the other hand on the decomposition into 8 degrees of freedom. To finish, we provide one more proof to support this model, by taking advantage of the fact that our system is a spin-1: we study the creation of tensorial arrangement of spin, which results in ellipticity noise in non-perturbative SNS.

\section{Higher-order spin arrangements : creation of ellipticity noise}
\subsection{Numerical simulation results}

\begin{figure}[h]
    \centering
    \includegraphics[width=0.45\columnwidth]{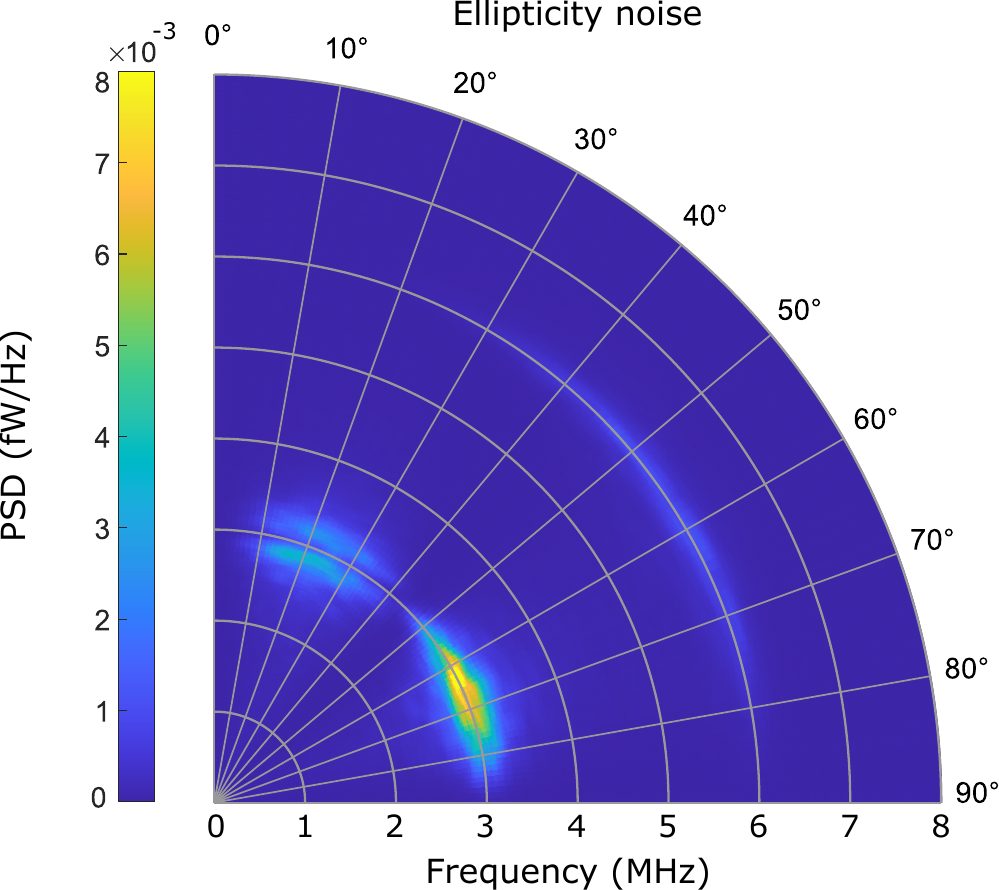}
    \caption{Ellipticity noise spectra obtained for $P = 1.5\,\mathrm{mW}$, $\Delta/2\pi = 1500\,\mathrm{MHz}$, and for $\theta$ varying from 0 to 90$^{\circ}$. The magnetic noise correlation time is $\tau_{\mathrm{c}} = 5.3\times 10^{-9}\,\mathrm{s}$, its standard deviation corresponds to 12\% of $\omega_{\mathrm{L}}$ with $\omega_{\mathrm{L}}/ 2\pi=3\,\mathrm{MHz}$. Each radius corresponds to one spectrum obtained for the corresponding value of $\theta$.}
    \label{simu_ellip}
\end{figure}

In a previous article \cite{PhysRevA.107.023527}, we discussed the ellipticity noise that can arise from higher-order tensorial arrangements. They correspond to linear birefringence noise created by alignment degrees of freedom \cite{fomin_spin-alignment_2020}. This noise is measured as an ellipticity noise when adding a quarter-wave plate (QWP) before the balanced detection (see \cite{fomin_spin-alignment_2020, Liu_2022, poltavtsev_spin_2014} for details). We know from these prior SNS experiments that ellipticity noise can be detected at both the Larmor frequency and twice the Larmor frequency. Indeed, in the first case, the degrees of freedom $M_{5,6}$ are involved, which oscillate at $\omega_\mathrm{L}$ and are responsible for linear birefringence with neutral axes oriented at $\theta = \pm 45^{\circ}$. As a consequence, no noise is observed near $\theta = \pm 45^{\circ}$. The noise at $2\omega_\mathrm{L}$ is created by noise in the degrees of freedom $M_4,\, M_7$, and $M_8$, responsible for linear birefringence with neutral axes oriented at $0$ and $ 90^{\circ}$. As such, the noise is not visible near $\theta = 0$ and $90^{\circ}$.  We now show that those tensorial arrangements can emerge from the noise in the transverse magnetic field, just as the FR noise studied above.

The simulated birefringence noise spectra can be seen in figure \ref{simu_ellip}, in the same conditions as in figure \ref{simu_rot}. The simulations show indeed that some noise is created at $\omega_{\mathrm{L}}$. As expected, this noise is not visible near $\theta = 45^{\circ}$. However, it is also zero near $0$ and $90^{\circ}$: this means again that the magnetic field fluctuations do not create spin noise when the probe beam is aligned in these directions, since it would necessarily be measured otherwise. One can also see that some noise is created near $2\omega_\mathrm{L}$, absent for $\theta = 0^{\circ}$ and $90^{\circ}$. Contrary to the previous case, this behavior is similar to the case where the spin noise originates from transit noise.

\subsection{Steady-state excitation of tensorial degree of freedom}
\begin{figure}
    \centering
    \includegraphics[width=0.55\columnwidth]{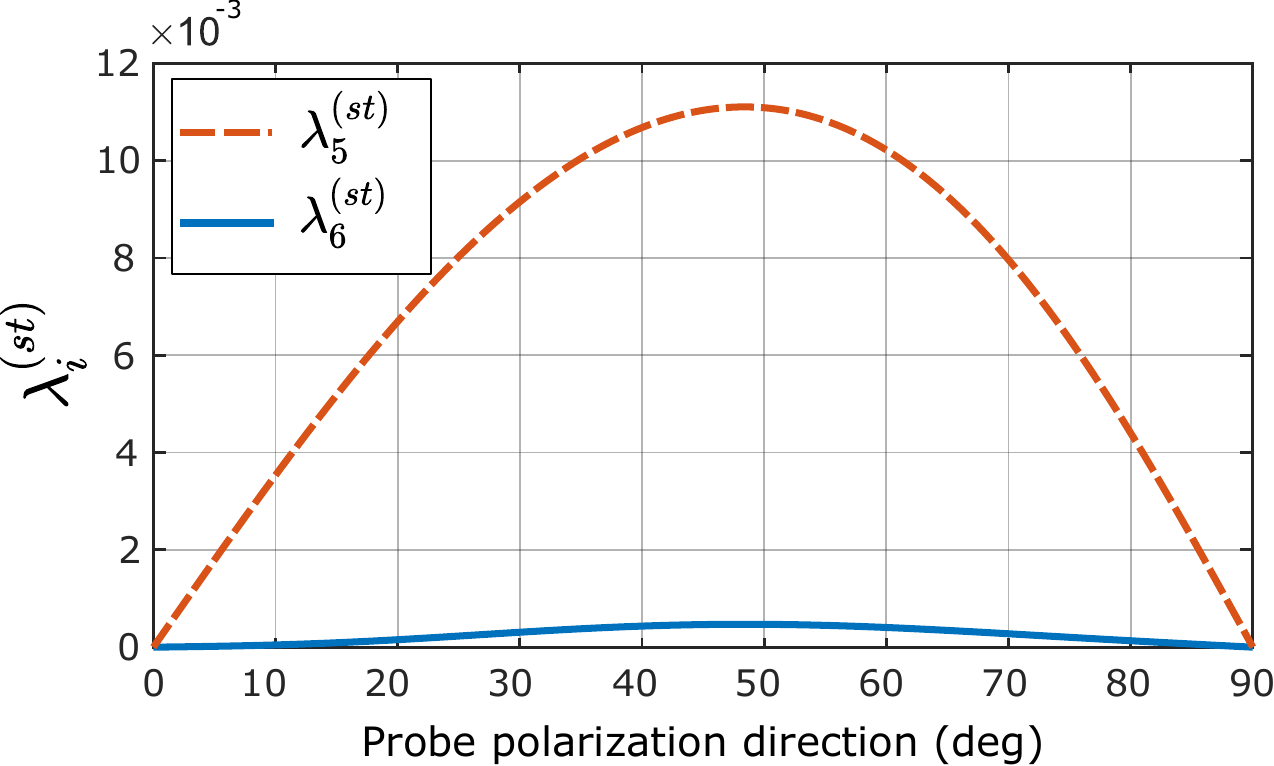}
    \caption{Evolution versus $\theta$ of the coefficients $\lambda_5^{(\mathrm{st})}$ and $\lambda_6^{(\mathrm{st})}$ obtained at steady-state. Same  parameters as in figure \ref{stationary_transverse_spin}.}
    \label{stationary_M5_6}
\end{figure}

The fact that the ellipticity noise at the Larmor frequency in figure  \ref{simu_ellip} is absent near $\theta = 0^{\circ}$ and $90^{\circ}$ can be explained in a similar manner as for the FR noise. Indeed, this ellipticity noise is attributed to the modes $M_5$ and $M_6$. These spin degrees of freedom are coupled to one another by the magnetic noise. Indeed, the commutator $\left[ M_2,\, {\mathrm{e}}^{-iM_2\omega_\mathrm{L}(t-t')}\,M_j\, {\mathrm{e}}^{iM_2\omega_\mathrm{L}(t-t')}\right]$  in (\ref{eq16}) maps this subset of matrices $\{ M_j\}_{j=5,6}$ to itself. However, our simulations of steady state show that none of these two modes is populated due to optical pumping for $\theta = 0^{\circ}$ and $90^{\circ}$. Indeed, figure \ref{stationary_M5_6} show both coefficients $\lambda_5^{(\mathrm{st})}$ and $\lambda_6^{(\mathrm{st})}$ at steady-state in the same conditions as above, clearly showing that both coefficients are zero for $\theta = 0^{\circ}$ and $90^{\circ}$. This leads to the absence of ellipticity noise at $\omega_\mathrm{L}$ in these light polarization directions.

Regarding the noise at $2\,\omega_\mathrm{L}$, the same analysis of the steady-state shows that there is always at least one of the degrees of freedom $M_4$, $M_7$ or $M_8$ which is excited for every value of the angle $\theta$. These matrices are coupled to one another by the above commutator, and are responsible for linear birefringence, creating ellipticity noise at $2\,\omega_\mathrm{L}$. As a consequence, this alignement noise is excited whatever the orientation of the polarization. The fact that this noise is not measured near $\theta = 0^{\circ}$ or $90^{\circ}$ comes from the fact that these directions are the neutral axes of the corresponding fluctuating linear birefringence.

\section{Conclusion}
We have theoretically investigated the effect of the presence of amplitude noise in the transverse magnetic field used in a spin noise spectroscopy experiment, in the case where the sample under study is a spin-1 system. We first introduced a theoretical model for a $J=1 \rightarrow J=0$ transition and implemented a numerical resolution scheme, which eventually allowed us to simulate spin noise spectra. Interestingly, FR noise was indeed observed by introducing only a magnetic field noise acting on an open system, without taking any stochastic fluctuations of the populations of the Zeeman sublevels into account. The variance of the simulated spin noise increases linearly with the magnetic noise power, and reaches levels potentially comparable with the ones measured experimentally and attributed to the transit of the atoms through the beam. We also conducted a polarization-resolved study, which revealed  unusual polarization dependence, exhibiting in particular some directions along which no noise could be measured. 

To provide a physical picture for these simulation results, we developed an analytical model based on a first order perturbative approximation of the spin equation of motion. Using the decomposition of the density matrix in 8 independent degrees of freedom, we have shown that the magnetic fluctuations disturb the steady-state of the system and randomly couple some of these degrees of freedom together. In particular, this leads to a stochastic precession of the spin around its steady-state, thus explaining the Faraday rotation noise. We analyzed the dependence of the spin noise on the magnetic noise bandwidth and modulation frequency, to show that the critical parameter regarding the relevant spectral components of the magnetic noise is the Larmor frequency. Therefore, only the noise power density in the frequency band containing the natural oscillation frequency $\omega_{\mathrm{L}}$ excites the spin noise.

We then highlighted the tremendous impact of the steady-state around which the fluctuations are induced. We have shown that the residual absorption of the probe is responsible for optical pumping, which can lead to the existence of a steady-state with a non-zero average spin. If this spin exhibits a component in the plane orthogonal to the magnetic field,  stochastic precession then can occur, and some spin noise is observed. On  the contrary, this component does no longer exist for some other directions of the probe polarization, when the spin is aligned with the magnetic field at steady-state: the fluctuations then do not create any spin noise. 

Finally, we have shown that such a magnetic noise could also be responsible for the fluctuations of higher-order spin arrangements in a spin-1 system. We simulated some ellipticity spectra, showing noise both at $\omega_\mathrm{L}$ and  $2\,\omega_\mathrm{L}$. We could again explain the specific polarization dependence by looking at the excitation of these higher-order degrees of freedom at steady-state. We were able to distinguish between the angles, for which no noise is created by the joint action of the optical pumping and of the magnetic field and the directions, for which no noise is measured simply because they correspond to the neutral axes of the corresponding linear birefringence noise.


These results highlight specific properties of this usually unwanted noise, that helps understanding fundamental decoherence processes due to the random driving of a system by external fields. Furthermore, the characterization of the spin noise versus the probe polarization power, as well as of the probe polarization dependence, could help subtracting or suppressing this additional spin noise in SNS experiments. On the other hand, our observation that the extra spin noise power depends linearly on magnetic noise variance, and is excited only by the magnetic power density at $\omega_\mathrm{L}$, could be used to (i) develop a high bandwidth, optical spectrum analyzer for ac magnetic field; (ii) design strategies in precision magnetometry where this spectral analysis allows for the stabilization of magnetic sources. In both cases, further investigations could be done to assess how the magnetic noise affects the quantum limit for the spin projection noise.

\begin{acknowledgments}
 The authors acknowledge funding by the Institut Universitaire de France, the Labex PALM and the Chinese Scolarship Council.
\end{acknowledgments}

\appendix
\section{Derivation of the perturbative equation of motion of the spin}

Starting from equation (\ref{eq10}), one gets rid of the precession due to the term $[\bar{H},\delta\rho]/i\hbar$ by moving to the frame rotating in a deterministic manner. We thus write $\delta\Tilde{\rho} = \exp \left(i\bar{H} t/\hbar\right) \,\delta\rho \,\exp \left(-i\bar{H} t /\hbar\right)$ to get the following equation of motion : 
\begin{equation}
    \frac{\mathrm{d}\delta\Tilde{\rho}}{\mathrm{d}t} -\frac{1}{i\hbar}{D'}(\delta\Tilde{\rho}) = \frac{1}{i\hbar}[\delta H(t),\Tilde{\rho}^{\mathrm{st}}]\ ,
    \label{eqA1}
\end{equation}
where $\Tilde {\rho}^{\mathrm{st}} $ now depends on time according to the change of frame.
In the following, we suppose that the action of the probe light impacts the steady-state $\rho^{(\mathrm{st})}$, but that the dynamics of $\delta \rho$ can be described by the magnetic field only, so that we reduce the dissipation $D'(\rho)$ to that undergone by the Zeeman lower sublevels. We take a relaxation rate $\gamma$ identical for population and coherences. Equation (\ref{eq11}) becomes:
\begin{equation}
    \frac{\mathrm{d}\delta\Tilde{\rho}}{\mathrm{d}t} +\gamma\,\delta\Tilde{\rho} = \frac{1}{i\hbar}[\delta H(t),\Tilde{\rho}^{\mathrm{st}}]\ .
\label{eqA2}
\end{equation}
This equation can be formally integrated to give:
\begin{equation}
    \delta \rho(t) = -\dfrac{i}{\hbar}\int_{-\infty}^t\mathrm{dt'} \,\left[ \delta H (t'),\, \Tilde{\rho}^{\mathrm{st}}\,\right] \exp^{-\gamma(t-t')}\ .
    \label{eqA3}
\end{equation}
Transforming back $\delta\Tilde{\rho}$ to the lab reference frame finally leads to the following expression for the time evolution of the fluctuations $\delta\rho$:

\begin{equation}
\delta \rho(t) = -\dfrac{i}{\hbar}\int_{-\infty}^t\mathrm{dt'} \,\left[ \delta H (t'),\, {\mathrm{e}}^{-i\bar{H}(t-t')/\hbar}\,\rho^{\mathrm{st}}\, {\mathrm{e}}^{i\bar{H}(t-t')/\hbar}\right] \exp^{-\gamma(t-t')}\ ,
    \label{eqA4}
\end{equation}
which is identical to equation (\ref{eq11}).

\section{Expansion of the spin-1 density matrix around steady-state}

In the context of spin noise where one observes the fluctuations of the system around its thermal equilibrium state, the density matrix of a spin one system can be written as the sum of two parts, namely the thermal equilibrium state and the surrounding fluctuations:
\begin{equation}
    \rho = \dfrac{1}{3} \mathbf{1}+\frac{1}{2}\sum_{i=1}^8 \lambda_{i} M_{i}\label{rhoFunction}\ .
\end{equation}
In this equation the fluctuations are expanded over the spin operators of a single particle $M_{i}$, $i=1..8$, with coefficients ${\lambda}_{i}\equiv  \mathrm{Tr}[\rho M_{i}]$.
The $M_{i}$'s are traceless Hermitian operators, which obey the orthogonality relations $\mathrm{Tr}(M_{i} M_{j})=2\delta_{ij}$.  With the quantization axis along $z$, i.e. in a basis consisting in the three kets $\{\ket{-1}_z,\ket{0}_z,\ket{1}_z\}$, the first three operators $M_1,\,M_2,\,M_3$ describe the polarization of the spin along the directions $z$, $x$, and $y$, respectively:
\begin{equation}
M_{1}=\left[\begin{array}{ccc}
    1&0&0  \\
     0&0&0\\
     0&0&-1
\end{array}\right]\ ,
\end{equation}
\begin{equation}
     M_{2}=\frac{1}{\sqrt{2}}\left[\begin{array}{ccc}
    0 &1&0  \\
     1&0&1\\
     0&1&0
    \end{array}\right]\ ,
\end{equation}
\begin{equation}
    M_{3}=\frac{1}{\sqrt{2}}\left[\begin{array}{ccc}
    0 &-i&0  \\
     i&0&-i\\
     0&i&0
     \end{array}\right]\ .
\end{equation}
The five remaining operators are represented by the following matrices: 
\begin{equation}
    M_{4}=\left[\begin{array}{ccc}
    0 &0&1  \\
     0&0&0\\
     1&0&0
\end{array}\right]\ ,
\quad M_{5}=\left[\begin{array}{ccc}
    0 &0&-i  \\
     0&0&0\\
     i&0&0
\end{array}\right]\ ,
\end{equation}
\begin{equation}
    M_{6}=\frac{1}{\sqrt{2}}\left[\begin{array}{ccc}
    0 &1&0  \\
     1&0&-1\\
     0&-1&0
\end{array}\right]\ ,
\quad M_{7}=\frac{1}{\sqrt{2}}\left[\begin{array}{ccc}
    0 &-i&0  \\
     i&0&i\\
     0&-i&0
\end{array}\right]\ ,
\end{equation}
\begin{equation}
   M_{8}=\frac{1}{\sqrt{3}}\left[\begin{array}{ccc}
         1&0&0  \\
         0&-2&0\\
         0&0&1
    \end{array}
    \right]\ .
\end{equation}
The operators $M_{4}, M_{5}$ describe coherences between the spin states $\ket{-1}_{z}$ and $\ket{+1}_z$, while $M_{6}$ and $M_{7}$ describe coherences between $\ket{0}_z$ and the states $\ket{\pm 1}_{z}$. Finally $M_{8}$ describes the spin alignment corresponding to population imbalance between $\ket{0}_z$ and the other two states. 

In section 3 of this paper, we rather decompose the state as the sum of  the steady-state and of the surrounding fluctuations. As such, one can distinguish between the coefficients $\lambda_i$ at steady-state, and their perturbation at  first order, i. e. $\lambda_i = \lambda_i^{\mathrm{st}} + \lambda_i^{(1)}$. Then (\ref{rhoFunction}) can be rewritten as
\begin{equation}
    \rho = \left(\frac{1}{3} \mathbf{1}+ \frac{1}{2}\sum_{i=1}^8 \lambda_{i}^{(\mathrm{st})} M_{i}\right)+ \frac{1}{2}\sum_{i=1}^8 \lambda_{i}^{(1)} M_{i} = \rho^{\mathrm{st}} + \frac{1}{2}\sum_{i=1}^8 \lambda_{i}^{(1)} M_{i}\ .
\end{equation}

\section{Integration of the equation of motion : computation of the spin correlator $\overline{\lambda_z^{(1)}(T) \lambda_z^{(1)}(0)}$}

Applying the decomposition of  (\ref{eq13}) to $\delta \rho$, equation (\ref{eq14})  shows that  a fluctuation of the degree of freedom $M_i$ \textit{around steady state} (i.e., a fluctuation of the coefficient $\lambda^{(1)}(t)$) can emerge from the other arrangements $M_j$ \textit{at steady-state} thanks to the the magnetic field noise provided the commutator $\left[ M_2,\, {\mathrm{e}}^{-iM_2\omega_\mathrm{L}(t-t')}\,M_j\, {\mathrm{e}}^{iM_2\omega_\mathrm{L}(t-t')}\right]$ effectively maps the operator $M_j$ to $M_i$. We will prove this statement by investigating the conventional Faraday rotation noise, which emerges from the fluctuations of $S_z$. Two matrices $M_i$ couple to $S_z = M_1$ by the above commutator : $M_1$ itself and $M_3 = S_y$ (the commutator of each couple ($M_i,\, M_j$) can be found in ref. \cite{colangelo_quantum_2013}). One can show that  (\ref{eq14}) gives the two coupled equations :

\begin{equation}
    \left \{
    \begin{array}{c c}
     \lambda_z^{(1)}(t) =& \lambda_y^{(\mathrm{st})} \int_{-\infty}^t \mathrm{dt'}\delta\omega_\mathrm{L}(t') \cos \omega_\mathrm{L} (t-t') {\mathrm{e}}^{-\gamma (t-t')}\ , \\
        &-\lambda_z^{(\mathrm{st})} \int_{-\infty}^t \mathrm{dt'} \delta\omega_\mathrm{L}(t') \sin \omega_\mathrm{L} (t-t') {\mathrm{e}}^{-\gamma (t-t')} \\
      & \\
     \lambda_y^{(1)}(t) =& -\lambda_z^{(\mathrm{st})} \int_{-\infty}^t \mathrm{dt'}\delta\omega_\mathrm{L}(t') \cos \omega_\mathrm{L} (t-t') {\mathrm{e}}^{-\gamma (t-t')} \\
     &-\lambda_y^{(\mathrm{st})} \int_{-\infty}^t \mathrm{dt'} \delta\omega_\mathrm{L}(t') \sin \omega_\mathrm{L} (t-t') {\mathrm{e}}^{-\gamma (t-t')}\ .
    \end{array} 
    \right.
    \label{eqC1}
\end{equation}

We then introduce two coefficients analog to the ladder operators: $\lambda_+ = \lambda_y+i\lambda_z$ and $\lambda_- = \lambda_y-i\lambda_z$. This greatly simplifies (\ref{eqC1}) which becomes : 

\begin{equation}
    \left \{
    \begin{array}{c c}
     \lambda_+^{(1)}(t) =& i \lambda_+^{(\mathrm{st})} \int_{-\infty}^t \mathrm{dt'} \,\delta \omega_\mathrm{L}(t') {\mathrm{e}}^{i\omega_\mathrm{L} (t-t') {\mathrm{e}}^{-\gamma(t-t')}}\ , \\ 
     & \\
     \lambda_-^{(1)}(t) =& -i\lambda_-^{(\mathrm{st})} \int_{-\infty}^t \mathrm{dt'} \,\delta \omega_\mathrm{L}(t') {\mathrm{e}}^{-i\omega_\mathrm{L} (t-t') {\mathrm{e}}^{-\gamma(t-t')}}\ .
    \end{array} 
    \right.
    \label{eqC2}
\end{equation}

Eventually, we want to compute the autocorrelation function for the $S_z$ component of the density matrix fluctuations. The coefficient $\lambda_z = \mathrm{Tr}\,\rho S_z = \langle S_z \rangle$ corresponds to the ensemble average (denoted by $\langle ... \rangle$) of the $z$-component of the spin, so that the corresponding spin noise autocorrelation function can be written $\overline{\lambda_z^{(1)}(T) \lambda_z^{(1)}(0)}$. To access it, one can compute the autocorrelation function $\overline{\lambda^{(1)}_+(T) \lambda^{(1)}_-(0)}$ and take its real part: $\mathrm{Re}\,[ \overline{\lambda^{(1)}_+(T) \lambda^{(1)}_-(0)} ]= \overline{\lambda_z^{(1)}(T) \lambda_z^{(1)}(0)} + \overline{\lambda_y^{(1)}(T) \lambda_y^{(1)}(0)} = 2\overline{\lambda_z^{(1)}(T) \lambda_z^{(1)}(0)}$, the last equality coming from the axial symmetry of the problem with respect to the $x$-axis.

We will consider the case studied numerically in section 1, i. e. that of a Gaussian Larmor frequency noise, with variance $\omega_\sigma^2$ and correlation time $\tau_{\mathrm{c}}$. We have :

\begin{equation}
    \overline{\lambda_+^{(1)}(T) \lambda_-^{(1)}(0)} = \lambda_+^{(\mathrm{st})}\lambda_-^{(\mathrm{st})} \int^T_{-\infty} d\tau_1 \int^0_{-\infty} d\tau_2\, \overline{\delta \omega_\mathrm{L} (\tau_1) \delta \omega_\mathrm{L}(\tau_2)} \,e^{i\omega_\mathrm{L} (T-\tau_1+\tau_2)}e^{-\gamma(T-\tau_1-\tau_2)}\ .
    \label{eqC3}
\end{equation}
After some algebra, one finds using (\ref{eq8})
\begin{align}
    \overline{\lambda_+^{(1)}(T) \lambda_-^{(1)}(0)} = \lambda_+^{(\mathrm{st})}\lambda_-^{(\mathrm{st})} \omega_\sigma^2\left[ \dfrac{{\mathrm{e}}^{-\gamma \vert T \vert +i\omega_\mathrm{L}T+i\phi_1}/\gamma\tau_{\mathrm{c}}}{\left[{\left( \gamma^2- \omega_\mathrm{L}^2-1/\tau_{\mathrm{c}}^2\right)^2+4\omega_\mathrm{L}^2\gamma^2}\right]^{1/2}}\right.\nonumber\\
    + \left.\dfrac{{\mathrm{e}}^{-\vert T\vert/\tau_{\mathrm{c}}+i\phi_2}}{\left[\left( \gamma^2+ \omega_\mathrm{L}^2-1/\tau_{\mathrm{c}}^2\right)^2+4\omega_\mathrm{L}^2/\tau_{\mathrm{c}}^2\right]^{1/2}}\right]\ ,
    \label{eqC4}
\end{align}
with $\tan \phi_1 = \varepsilon 2\omega_\mathrm{L}\gamma/(\gamma^2-\omega_L^2-1/\tau_{\mathrm{c}}^2)$ and  $\tan \phi_2 = \varepsilon 2\omega_\mathrm{L}/\tau_{\mathrm{c}}(\gamma^2+\omega_L^2-1/\tau_{\mathrm{c}}^2)$, with $\varepsilon =+1$ if $T>0$ and $-1$ if $T<0$.

In view of the complexity of such a correlator, we  make the assumption that the spin relaxation rate $\gamma$ is much smaller than both the Larmor frequency and the bandwidth of the magnetic noise: $\gamma \ll \omega_\mathrm{L}, \, 1/\tau_{\mathrm{c}}$. In that case, the first term in (\ref{eqC3}) dominates and the phase $\phi_1$ vanishes. One finally obtains the analytical expression (\ref{eq16}) for the fluctuation of $\lambda_z^{(1)}(t)$:

\begin{equation}
    \overline{\lambda_z^{(1)}(T) \lambda_z^{(1)}(0)} = \lambda_+^{(\mathrm{st})}\lambda_-^{(\mathrm{st})} \dfrac{\omega_\sigma^2\tau_{\mathrm{c}}}{2\gamma} \dfrac{1}{1+\omega_{\mathrm{L}}^2\tau_{\mathrm{c}}^2} \cos(\omega_{\mathrm{L}} T){\mathrm{e}}^{-\gamma\vert T\vert}\ .
    \label{eqC5}
\end{equation}

\section*{References}
\bibliography{bibliography.bib}

\end{document}